\documentclass[a4paper]{jpconf}
\usepackage{graphicx}

\usepackage{tikz}
\usepackage{xcolor}
\usepackage{psfrag,pstricks}
\usepackage{amsmath}
\usetikzlibrary{arrows}
\newcommand{\degree}{\ensuremath{^\circ}}
\colorlet{darkgreen}{green!60!black}

\pgfdeclarelayer{foreground}
\pgfdeclarelayer{background}
\pgfsetlayers{background,main,foreground}
\usepackage{url}
\usepackage{cite}

\usepackage[colorlinks=true,
            linkcolor=black,
            urlcolor=black,
            citecolor=black]{hyperref}

\begin{document}
\title{Improved energy production of multi-rotor wind farms}

\author{M. P. van der Laan$^1$ and M. Abkar$^2$}

\address{$^1$Technical University of Denmark, DTU Wind Energy, Ris\o \ Campus, DK-4000 Roskilde, Denmark\\
$^2$Aarhus University, Department of Engineering, DK-8000 Aarhus C, Denmark}

\ead{plaa@dtu.dk}

\begin{abstract}
The multi-rotor (MR) wind turbine concept can be used to upscale wind turbines without increasing the rotor diameter, which can be beneficial for transport, manufacture and design of wind turbines blades.
The rotor interaction of a MR wind turbine leads to a faster wake recovery compared to an equivalent single-rotor (SR) wind turbine wake.
In this article, the benefit of the faster wake recovery of MR wind turbines is quantified using Reynolds-averaged Navier-Stokes simulations of a 4$\times$4 rectangular MR wind farm, for three different inter wind turbine spacings.
The simulations predict an increase of 0.3-1.7\% in annual energy production for the MR wind farm with respect to an equivalent SR wind farm, where the highest gain is obtained for the tightest inter wind turbine spacing.
The gain in AEP is mainly caused by the aligned wind directions for the first downstream wind turbine in a wind turbine row of the MR wind farm, which is verified by an additional large-eddy simulation.
\end{abstract}

\section{Introduction}
The upscaling of wind turbine blades has led to multi-megawatt turbines equipped with blades approaching 100 m, which are challenging to design, transport and manufacture.
These challenges can be circumvented by the multi-rotor (MR) wind turbine concept, as discussed by Jamieson et al. \cite{Jamieson14}.
Between April 2016 and December 2018, Vestas Wind Systems A/S has built and operated a MR wind turbine demonstrator consisting of four 225 kW rotors and van der Laan et al. \cite{Laan19} have measured and modeled the corresponding power curve and velocity wake deficit.
Power curve measurements, Reynolds-averaged Navier-Stokes (RANS) simulations and multi-fidelity vortex model simulations showed a power increase of 0-2\% below rated due to the rotor interaction, which can lead to an increase in annual energy production (AEP) of about 1.5\%.
RANS simulations of the MR wind turbine demonstrator showed that the wake recovery distance is 1.04-1.44 $D_{\rm eq}$ shorter compared to a single-rotor (SR) wind turbine with an equivalent rotor diameter $D_{\rm eq}$, which can potentially lead to a tighter spaced wind farm layout.
Ghaisas et el. \cite{Ghaisas18} employed large-eddy simulations (LES) to show that a row of five MR wind turbines with an inter spacing of $4D_{\rm eq}$ has a reduced power deficit compared to a row of equivalent SR wind turbines when the wind direction is aligned with the row orientation.
In this work, the benefit of MR wind turbines in wind farms is further investigated in terms of AEP using RANS simulations of $4\times4$ 10 MW wind turbines for three inter spacings: 3, 4 and $5D_{\rm eq}$.
The simulation methodology and the results are discussed in Sections \ref{sec:method} and \ref{sec:results}, respectively.
The main RANS results are verified by an additional LES test case, as described in \ref{sec:AdditionalLEScase}.

\section{Methodology}\label{sec:method}
\begin{figure}[!h]
\centering
\includegraphics[scale=0.9,clip=true,trim=0 10 0 0]{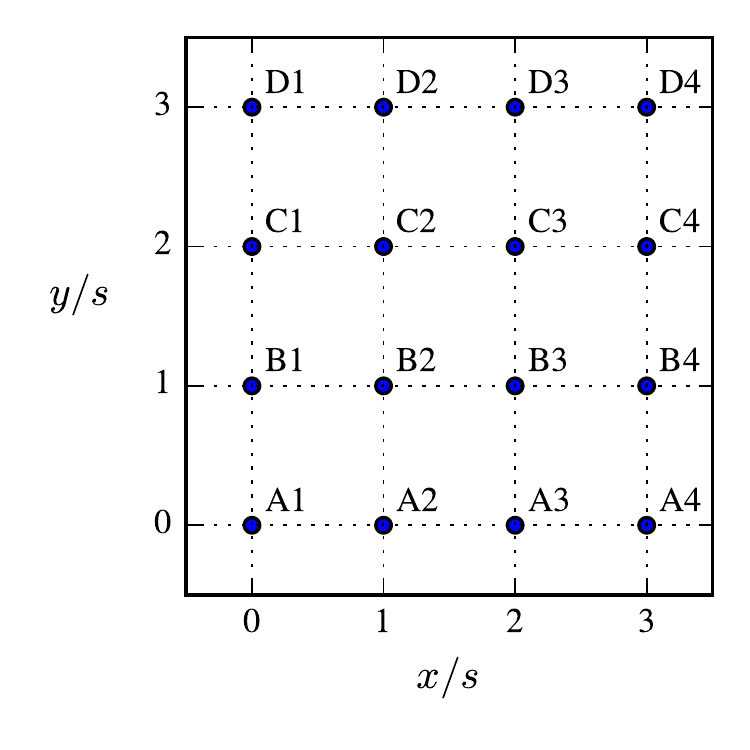}
\caption{Wind farm layout. Each point represents the wind turbine center.}
\label{fig:WFlayout}
\end{figure}

\begin{figure}[!h]
\centering
\includegraphics[scale=0.3,clip=true,trim=100 480 200 1000]{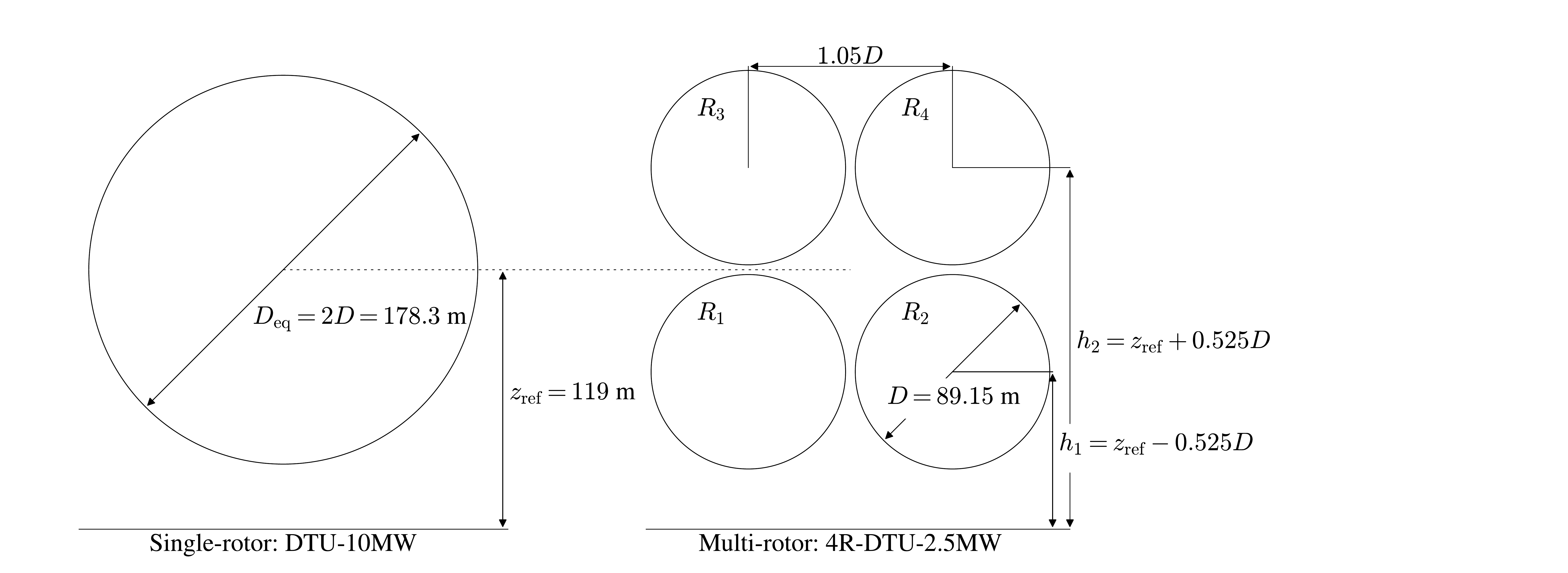}
\caption{Definition of SR and MR wind turbines.}
\label{fig:TurbineDef}
\end{figure}

\begin{figure}[!h]
\centering
\includegraphics[scale=0.7,clip=true,trim=0 10 0 0]{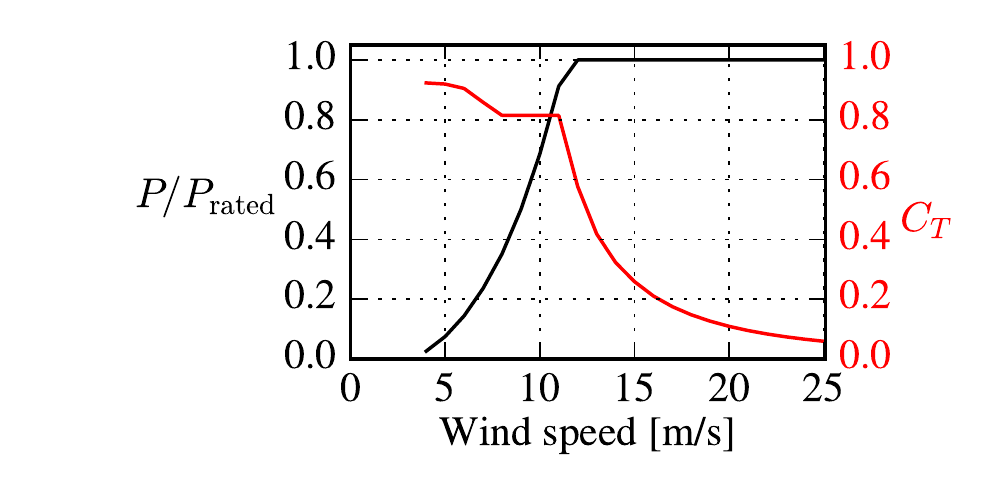}
\caption{Power and $C_T$ curves of the DTU-10MW wind turbine.}
\label{fig:Pct}
\end{figure}

RANS simulations of a $4\times4$ rectangular wind farm layout, as depicted in Figure~\ref{fig:WFlayout}, consisting of 10 MW wind turbines are carried out, for three inter spacings, $s$; 3, 4 and $5D_{\rm eq}$, and two different wind turbine types: a SR wind turbine and a MR wind turbine consisting of four rotors.
The SR wind turbine is based on the (offshore) DTU-10MW reference wind turbine \cite{DTU10MW}, which has a rotor diameter ($D_{\rm eq}$) and hub height ($z_{\rm ref}$) of 178.3 and 119 m, respectively.
The MR wind turbine consist of four rotors, where each rotor corresponds to a down scaled version of the DTU-10MW reference wind turbine with a rotor diameter ($D$) of 178.3/2=89.15 m.
Hence, the total swept area of the MR wind turbine is the same as the SR wind turbine.
The SR and MR wind turbine definitions are depicted in Figure~\ref{fig:TurbineDef}.
The center of the four rotors is located at the hub height of the DTU-10MW wind turbine (119 m). 
The horizontal and vertical spacing between the four rotors are taken as 5\% of the rotor diameter (4.46 m), which results in a lower and upper hub height of $z_{\rm ref}\pm0.525D$ ($h_1=72.20$ and $h_2=165.80$ m).
The four rotors of the MR wind turbine are oriented in the same plane, and there is no rotor tilt for both turbines.
Both the SR and MR wind turbines follow the power and thrust coefficient ($C_T$) curves as depicted in Figure~\ref{fig:Pct}.

The wind turbines are modeled as actuator disks (AD) \cite{MikkelsenRPhD,ADval} without rotation.
The effect of wake rotation on the power deficit is rather small, as shown in previous work \cite{Laan14c}, and by neglecting it the range of wind directions necessary to calculate the AEP reduces to a sector of 45\degree\ (for example between 270--315\degree) because of the chosen rectangular wind farm layout.
A wind direction and wind speed interval of 3\degree\ and 1 m/s is chosen, respectively, which results in 16 wind directions and 13 wind speeds (between 4-16 m/s) that needs to be simulated (208 cases per wind turbine type and inter spacing).

The numerical setup of the RANS simulations is described in detail in previous work \cite{Laan13b,Laan14b,Laan14c} and a summary is presented here.
The RANS equations are solved with EllipSys3D, a finite volume based flow solver initially developed by S{\o}rensen \cite{NielsPHD} and Michelsen \cite{Basis3D}.
The QUICK scheme \cite{QUICK} is used to discretize the convective terms and a SIMPLE algorithm \cite{SIMPLE} is employed to solve the pressure correction equation.
A modified Rhie-Chow algorithm \cite{RhieChowMod,ADNielsRev} is used to avoid pressure-velocity decoupling when using body forces.
The Cartesian flow domain has dimensions of about $1000D\times1000D\times50D$ in the streamwise, lateral and vertical directions, respectively.
The wind farm is located at the center, where a uniform spaced area in the horizontal direction is used to model the wind turbine wakes.
The refined inner domain is large enough to include the wind farm layout plus $6D$ upstream, $24D$ downstream and $6D$ lateral margins.
In previous work \cite{Laan19}, a cell spacing of $D/20$ was used to capture the 0-2\% power increase due to the rotor interaction of the MR wind turbine.
In the present work, it is shown in Section \ref{sec:ggridstudy} that a cell spacing of $D/8$ is fine enough to capture the MR wake deficit, and this spacing will be applied in the MR wind farm simulations.
The SR wind farm simulations are applied with the same numerical grid.
The flow direction is always set to 270\degree, and different wind direction are simulated by rotating the wind farm layout.
This avoids additional numerical diffusion that occurs when the wind direction is not aligned with the grid lines.
The bottom wall is set as a rough wall boundary condition \cite{nsqrBC}, the west and top boundaries are inlet conditions, at which a neutral logarithmic surface layer is applied, the east boundary is an outlet, at which the normal gradients are assumed to zero, and the lateral boundary conditions are periodic.
The applied neutral atmospheric surface layer is defined by a turbulence intensity of $I_{\rm ref}=\sqrt{2/3 k}/U_{\rm ref}=0.056$ at the reference height $z_{\rm ref}=119$ m ($k$ is the turbulent kinetic energy, and $U_{\rm ref}$ is the reference velocity, at $z_{\rm ref}$).
This gives a roughness length of about $10^{-4}$ m, which corresponds to an offshore roughness.

The turbulence is modeled by the $k$-$\varepsilon$-$f_P$ model \cite{Laan13b}, developed to simulate wind turbines wakes in an atmospheric surface layer.
The $k$-$\varepsilon$-$f_P$ is a modified $k$-$\varepsilon$ model that uses a local turbulence length scale limiter to improve the near wake velocity deficit \cite{Laan18a}.

\subsection{Improved variable AD force method suitable for low wind speeds}
\begin{figure}[!h]
\centering
\includegraphics[scale=0.7,clip=true,trim=0 20 0 0]{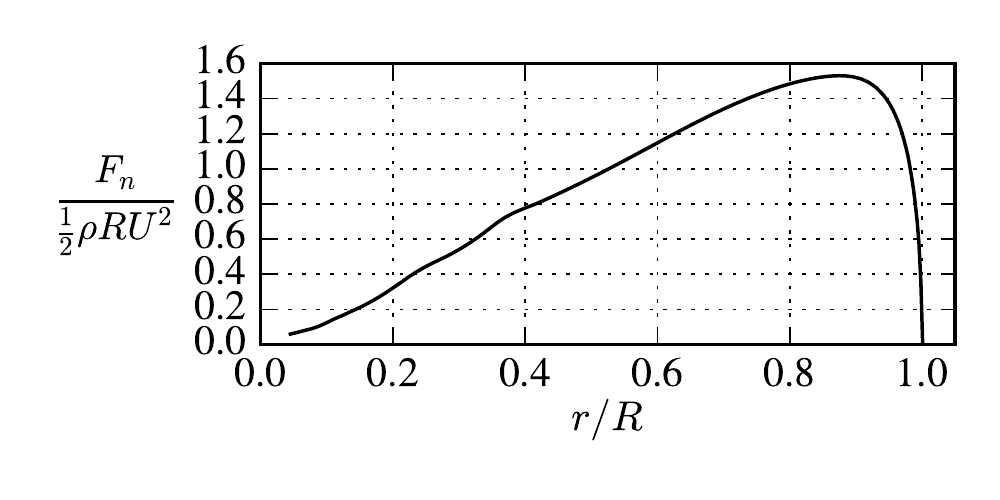}
\caption{Thrust force distribution $F_n$ of the DTU-10MW reference wind turbine for 8 m/s, obtained by a rotor-resolved detached-eddy simulation.}
\label{fig:Fnref}
\end{figure}

In previous work \cite{Laan14b}, a variable AD force method has been developed, where a normalized thrust force distribution, as depicted in Figure~\ref{fig:Fnref}, is scaled by the local AD velocity averaged over the AD: $\langle U_{\rm AD}\rangle$, and an alternative thrust coefficient $C_T^*$ that is a function of $\langle U_{\rm AD}\rangle$. 
$C_T^*$ can be calculated as:
\begin{align}
\label{eq:Ctstar}
C_T^*=C_T \left(\frac{U_{\rm ref}}{\langle U_{\rm AD}\rangle}\right)^2    
\end{align}
where $C_T$ is the standard thrust coefficient based on the freestream velocity $U_{\rm ref}$.
The $C_T^*$-$\langle U_{\rm AD}\rangle$ relation is obtained by single AD simulations for wind speeds between cut-in and cut-out with 1 m/s interval.
The variable force method works well for wind speeds above the cut-in the wind speed \cite{Laan14b,Laan14c}; however, for ADs operating around the cut-in wind speed (typically occurs when calculating the AEP), the method can cause numerical problems because two solutions can exist.
For example, for $U_{\rm ref}=3.99$ m/s, we can get that the AD force method alternates between on (non zero force) and off (zero force) for consecutive iterations.
One could include a (0,0) point in the $C_T^*$-$\langle U_{\rm AD}\rangle$ curve to avoid this issue, but this does not provide realistic results below the cut-in wind speed.
In this work, a better solution is proposed by using an additional on/off switch, where the AD is turned off if the estimated freestream velocity is lower than 95\% of the cut-in wind speed. 
The freestream velocity estimate is based on 1D momentum theory:
\begin{align}
U_{\rm ref}^{\rm 1D mom}=\langle U_{\rm AD}\rangle \left(1+\frac{C_T^*}{4}\right)    
\end{align}
by substitution of  $C_T=4a(1-a)$ and $a=1-\langle U_{\rm AD}\rangle/U_{\rm ref}$ in equation (\ref{eq:Ctstar}), where $a$ is axial induction.
$U_{\rm ref}^{\rm 1D mom}$ is only used as on/off switch, while the AD forces are based on the $C_T^*$-$\langle U_{\rm AD}\rangle$ relation.

Each rotor in the MR wind turbine is controlled separately using its own $C_T^*$-$\langle U_{\rm AD}\rangle$ relation. 
In practice, there is only a difference in the $C_T^*$-$\langle U_{\rm AD}\rangle$ curve between the bottom and top rotor pairs due to the wind shear.
By using calibrated $C_T^*$-$\langle U_{\rm AD}\rangle$ curves, free standing SR and MR wind turbines produce the same power, and the MR power gain due to the rotor interaction that was modeled in previous work using an AD force method based on airfoil data \cite{Laan19} is not taken into account in this article.
Hence, the difference between the SR and MR wind farm simulations is only caused by a difference in wake interaction.
If one would like to include both the power gain due to the rotor interaction and the faster wake recovery, then it is necessary to use the AD force method based on airfoil data together with a finer grid cell spacing around the ADs of $D/20$ (instead of $D/8$), as shown by grid refinement study performed in previous work \cite{Pirrung18}.
This would make the present RANS simulations about 20 times more expensive in terms of computational resources; therefore, only the power gain due to the faster wake recovery is investigated.



\section{Results and Discussion}\label{sec:results}
\subsection{Grid refinement study}\label{sec:ggridstudy}
\begin{figure}[!h]
\centering
\includegraphics[scale=0.8,clip=true,trim=0 10 0 5]{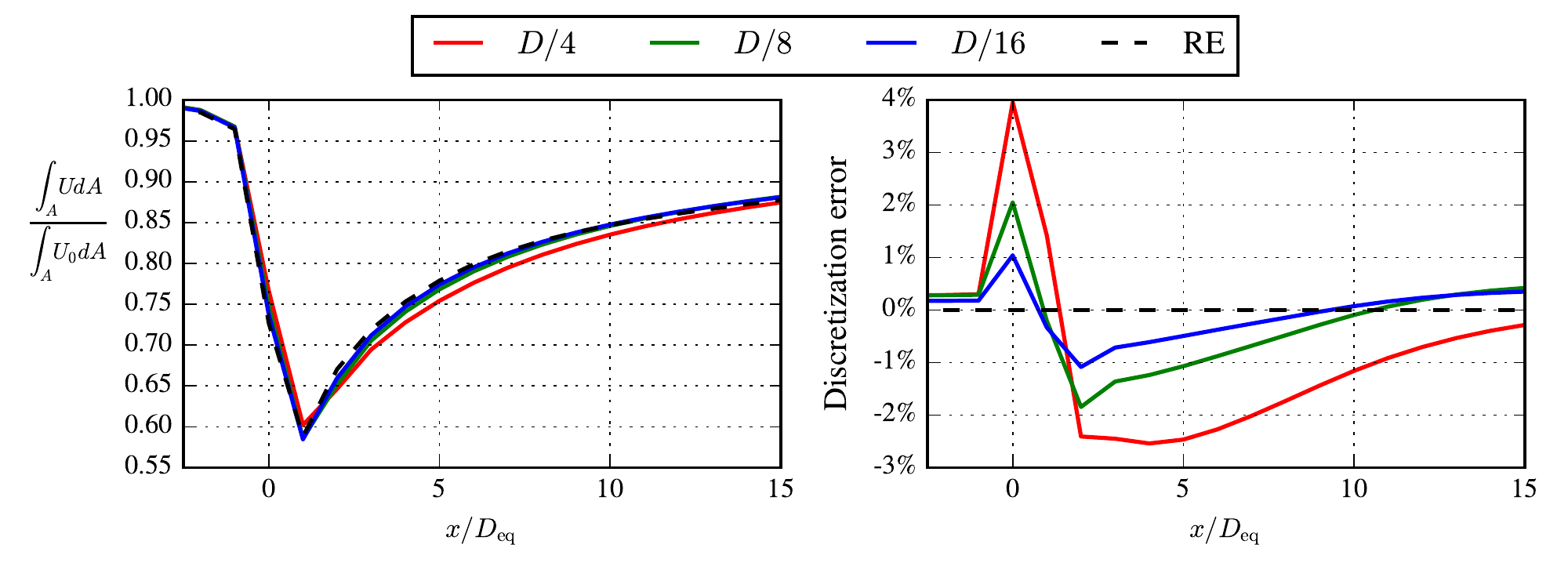}
\caption{Grid refinement study using three different grid sizes. Left: stream-wise velocity deficit integrated over the four rotor areas at different downstream distances. Right: Discretization error based on a mixed order analysis. RE = Richardson extrapolated value.}
\label{fig:GS}
\end{figure}

Results of a grid refinement study of the MR wind turbine are depicted in Figure~\ref{fig:GS}, where three different grid spacing are used in the domain around the MR: $D/4$, $D/8$ and $D/16$.
The numerical domain is similar to the wind farm simulations as described in Section \ref{sec:method}, but the size of the refined area around the wind turbine is smaller: $46D\times6D\times4D$ (the wind turbine center is located at $7D$ and $3D$ in stream-wise and lateral directions from the start of the refined area).
The stream-wise velocity deficit integrated over all four rotor areas is plotted as function of the downstream distance in the left plot of Figure~\ref{fig:GS}, for the three grid spacings, and an Richardson extrapolated value, calculated by a mixed-order analysis \cite{mixedOrder,ADval}.
The downstream integrals represents the wind speed that is available for a fictitious downstream MR wind turbine that is aligned with the modeled MR wind turbine.
The right plot of Figure~\ref{fig:GS} shows that the estimated discretization error (also calculated by the mixed-order analysis) is less than 2\% for a grid spacing of $D/8$ (and less than 1.4\% for $x\leq3D_{\rm eq}$, which is relevant for the MR wind farm using the smallest spacing of $s=3D_{\rm eq}$).
Hence, a grid spacing of $D/8$ is fine enough to perform wind farm simulations of MR wind turbines.
In addition, a finer grid spacing ($D/16$) leads to a smaller deficit than the chosen grid size ($D/8$). 
This makes the use $D/8$ a conservative choice when comparing the MR wind farm simulations with SR wind farm simulations because the same numerical grid is used for both wind turbine types (leading to a grid size of $D_{rm eq}/16$ for the SR wind farm).

\subsection{SR wind farm compared MR wind farm}
In the proceeding sections, the difference between the SR and MR wind farms are discussed in term of velocity deficit, power deficit and AEP.

\subsubsection{Velocity deficit}\label{sec:ResVelDef}
\begin{figure}[!h]
\centering
\includegraphics[scale=1,clip=true,trim=10 175 0 20]{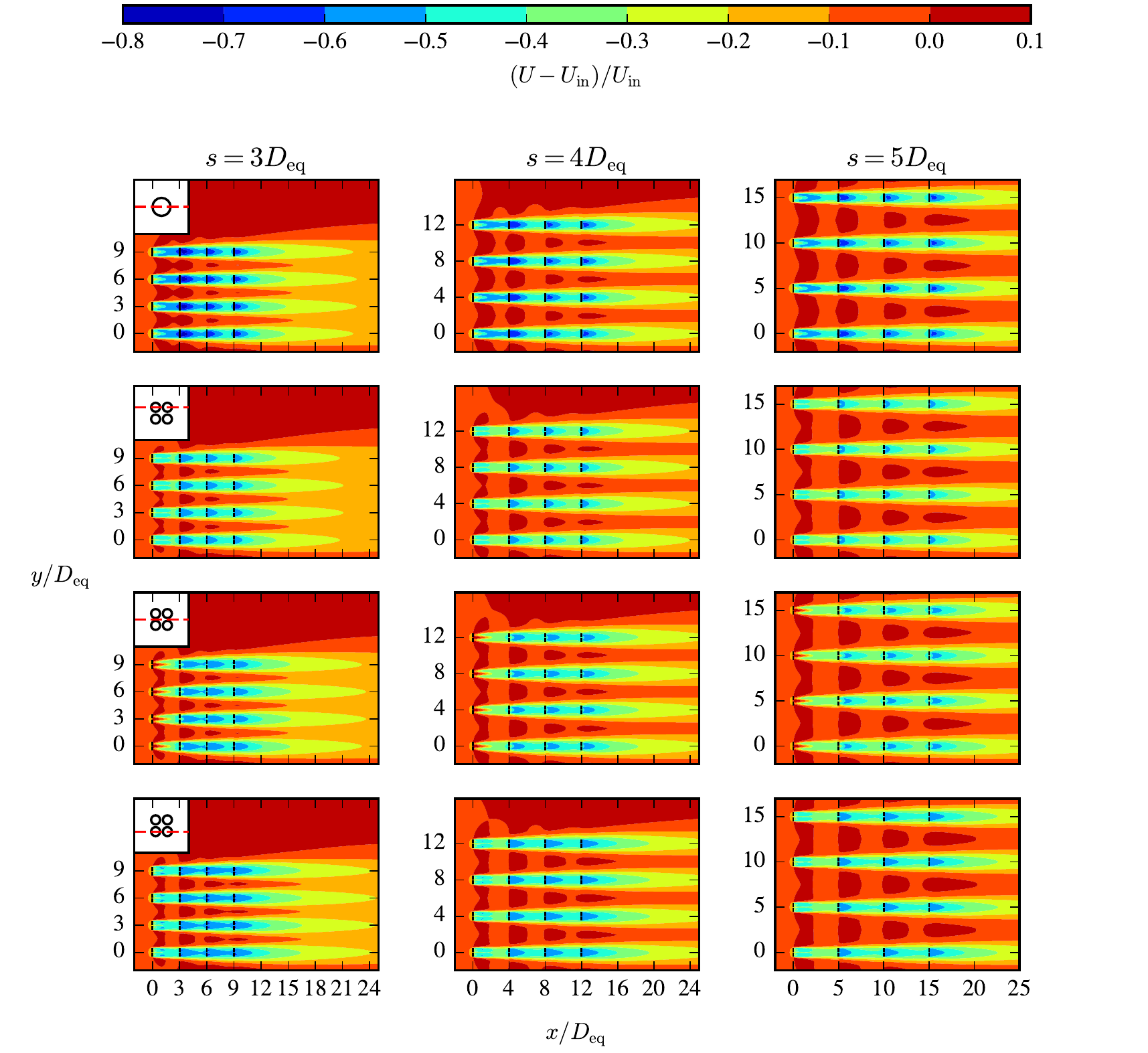}
\caption{Contours of stream-wise velocity normalized by the inflow profile, for the SR wind farm at $z=z_{\rm ref}$ (top row plots) and MR wind farm at $z=h_2$ (second row plots), $z=z_{\rm ref}$ (third row plots) and $z=h_1$ (bottom row plots), for three different inter spacings, for a reference velocity and wind direction of 8 m/s and 270\degree,\ respectively.}
\label{fig:VelDef}
\end{figure}
The velocity deficit of the SR and MR wind farms are depicted in Figure~\ref{fig:VelDef}, for three different inter spacings ($s/D_{\rm eq}=3$, 4 and 5), for a reference velocity and wind direction of 8 m/s and 270\degree,\ respectively. The velocity deficit plots of the SR wind farms are shown at hub height, which also corresponds to the reference height ($z=z_{\rm ref}$), while three heights are shown for the MR wind farm plots: the reference height ($z=z_{\rm ref}$), the hub height of the lower rotor pair ($z=h_1$), and the hub height of the top rotor pair ($z=h_2$).
The SR wind farm (top row) plots in Figure~\ref{fig:VelDef} show the typical near wake velocity deficits for the first free standing wind turbines (wind turbines A1, B1, C1 and D1), while more mixed velocity deficits are seen for the downstream wind turbines due to the wake generated turbulence.
The velocity deficits of the free standing wind turbines of the MR wind farms, at the lower and top hub heights (bottom and second rows of plots in Figure~\ref{fig:VelDef}) show distinct small SR velocity deficits, while the velocity deficits of the downstream wind turbines (e.g. wind turbines A2, A3 and A4) resemble merged wakes that are comparable with SR wind farm velocity deficits for downstream wind turbines.
This indicates that the largest difference in inflow wind speed due to wake effects between the SR and MR wind farms are observed for the second wind turbine in each row (wind turbines A2, B2, C2 and D2), while further downstream the difference in inflow wind speed is not large.
Two reasons could cause this effect; 1) The inflow turbulence is higher for the downstream wind turbines due to the upstream wind turbine wake, which decreases the difference between a SR and a MR wind turbine wake, as shown in previous work \cite{Laan19}; 2) The thrust force of the second wind turbine in a row is lower in the SR wind farm compared to the MR wind farm because of the lower inflow wind speed at this wind turbine in the SR wind farm, which reduces the difference in velocity deficit between the SR and MR wind farms for the third wind turbine in a row.

\subsubsection{Power deficit}\label{sec:ResPowerDef}
\begin{figure}[!h]
\centering
\includegraphics[scale=0.8,clip=true,trim=0 10 0 0]{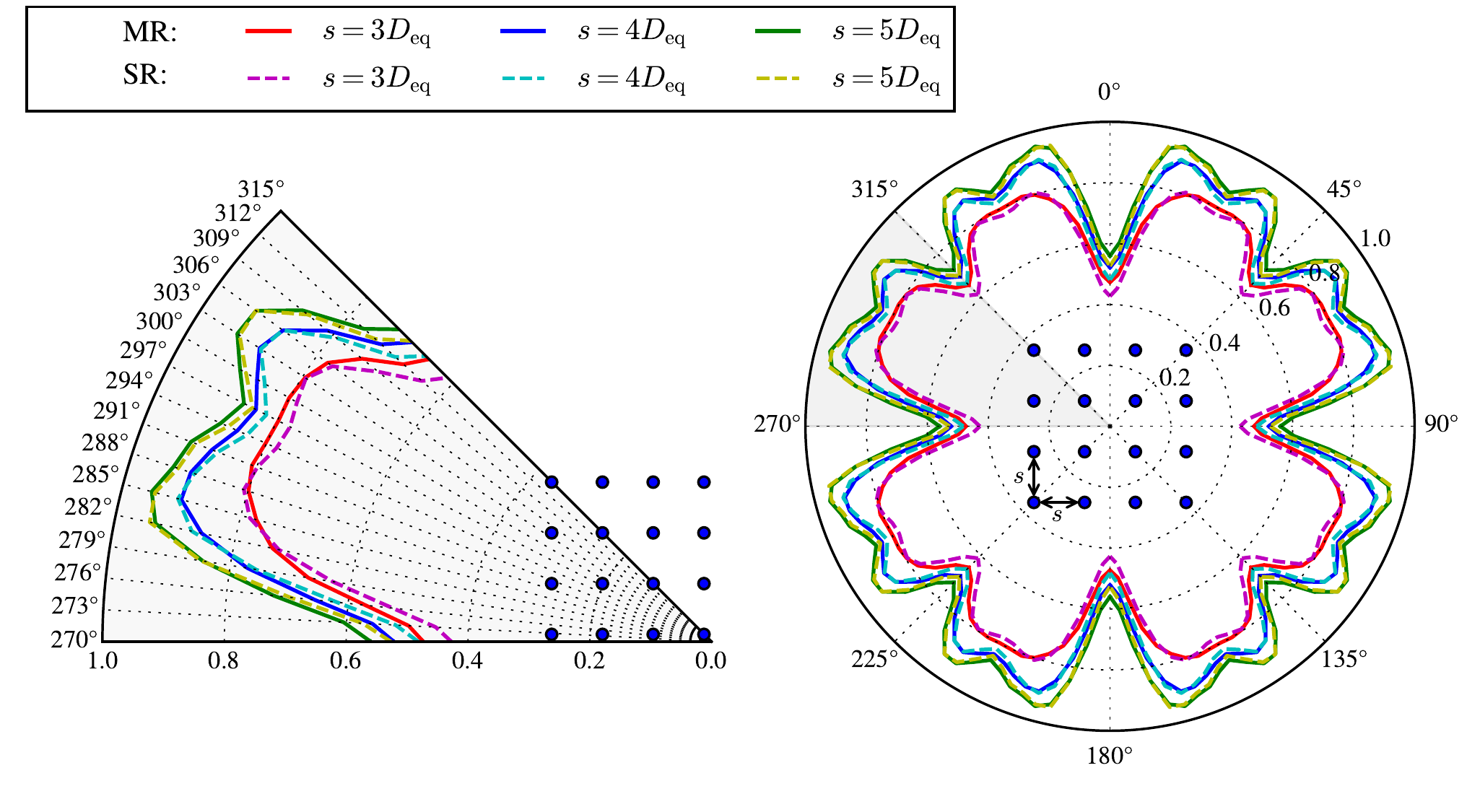}
\caption{Wind farm efficiency as functions of wind direction, for the SR and MR wind farms, for three different inter spacings, for a wind speed of 8 m/s. Left plot is a zoom of the right plot and represents the simulated wind directions. Wind farm layout is shown with blue dots.}
\label{fig:WFeff}
\end{figure}

The wind farm efficiency of the SR and MR wind farms are depicted in Figure~\ref{fig:WFeff}, for a wind speed of 8 m/s, and the three different inter spacings ($s/D_{\rm eq}=3$, 4 and 5).
The efficiency is calculated by normalizing the simulated wind farm power by the wind farm power without wake effects obtained by the power curve.
The actual simulated wind directions are shown in the left plot of Figure~\ref{fig:WFeff}, while the right plot depicts the wind farm efficiency for the entire wind rose using the symmetry of the wind farm layout.
Around the simulated aligned wind directions (mainly around 270 and 315\degree, but also for 297\degree), the MR wind farms have a higher efficiency than SR wind farms.
For 270 and 315\degree, the wind farm efficiency of the MR wind farm for $s=3D_{\rm eq}$ and  $s=4D_{\rm eq}$ corresponds to the wind farm efficiency of the SR wind farm for $s=4D_{\rm eq}$ and $s=5D_{\rm eq}$, respectively.
This finding corresponds with previous work \cite{Laan19}, where the wake recovery distance of the MR wind turbines was found to be 1.04-1.44 $D_{\rm eq}$ shorter compared to a SR wind turbine.
However, the observation is not true for the other wind directions, where the downstream wind turbines do not operate in full-wake inflow conditions.

\begin{figure}[!h]
\centering
\includegraphics[scale=0.8,clip=true,trim=0 10 0 0]{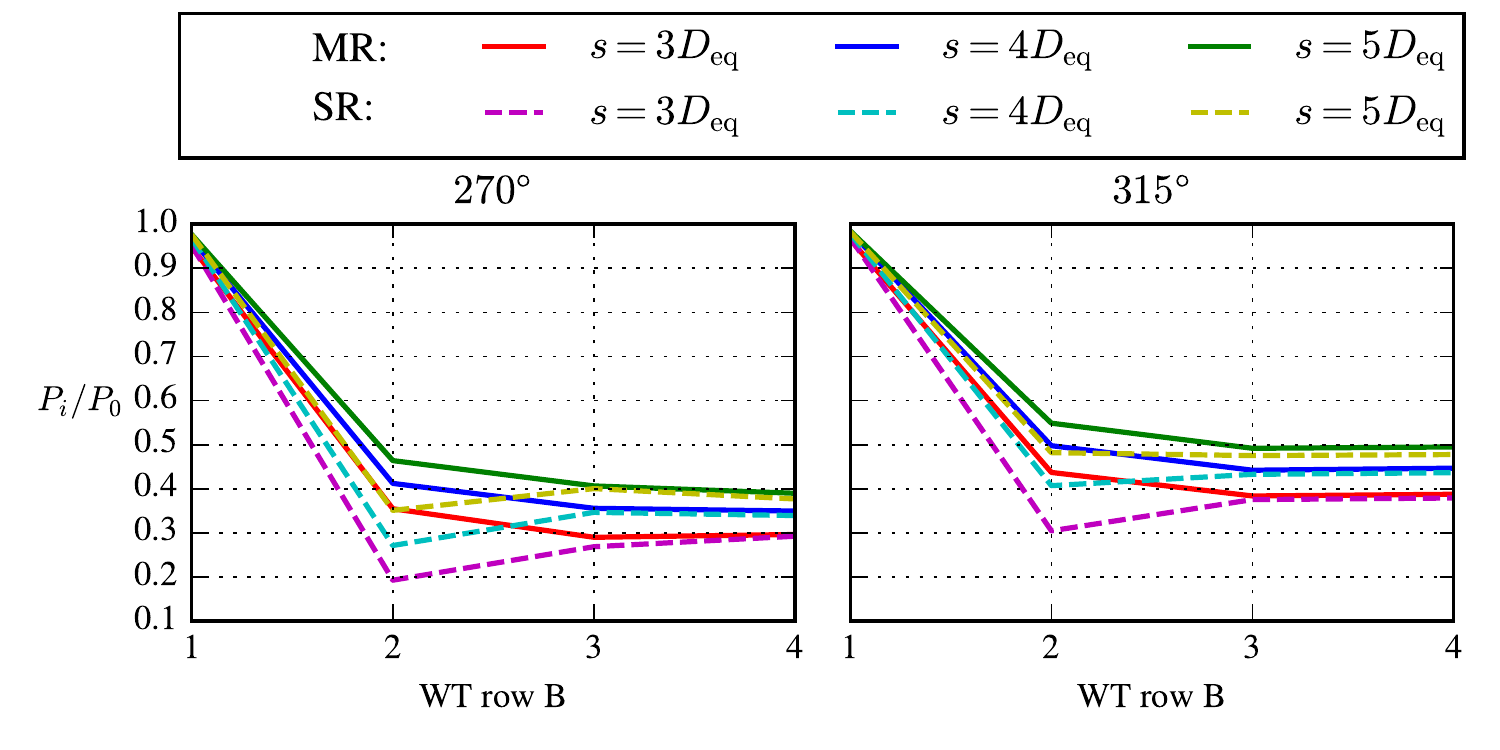}
\caption{Wind turbine (WT) power in row B, normalized by the wind turbine power for a single wind turbine. Results are shown for the SR and MR wind farms, for three different inter spacings, for a wind speed of 8 m/s. Left plot: aligned wind direction of 270\degree, right plot: misaligned wind direction of 282\degree.}
\label{fig:WFrow}
\end{figure}
The power deficit for wind turbine row B is shown in Figure~\ref{fig:WFrow} for the SR and MR wind farms, for the three investigated inter spacings, and for the two aligned wind directions.
Figure~\ref{fig:WFrow} confirms that the largest difference in power between the SR and MR wind farms is simulated for the second wind turbine in a row (11-16\% and 7-13\%, for a wind direction of 270 and 315\degree, respectively), as also discussed in Section \ref{sec:ResVelDef} for the velocity deficit.
The power deficit further downstream (wind turbines B3 and B4) is similar in both the SR and MR wind farms (difference is less than 2\%).
These results are significantly different from Ghasais et al. \cite{Ghaisas18}, who used LES to show that all downstream MR wind turbines in a row experience between 15-45\% smaller power deficits compared to a row of SR wind turbines, for an aligned wind direction.
It should be noted that Ghasais et al. \cite{Ghaisas18} used an unrealistically large tip clearance of $1D$, while we have used a more common tip clearance of $0.05D$ \cite{Laan18a}, as depicted in Figure~\ref{fig:TurbineDef}.
To investigate the results of Ghasais et al. \cite{Ghaisas18}, we have performed an additional test case based on LES, as described in \ref{sec:AdditionalLEScase}.
\begin{figure}[!h]
\centering
\includegraphics[scale=0.8,clip=true,trim=0 10 0 0]{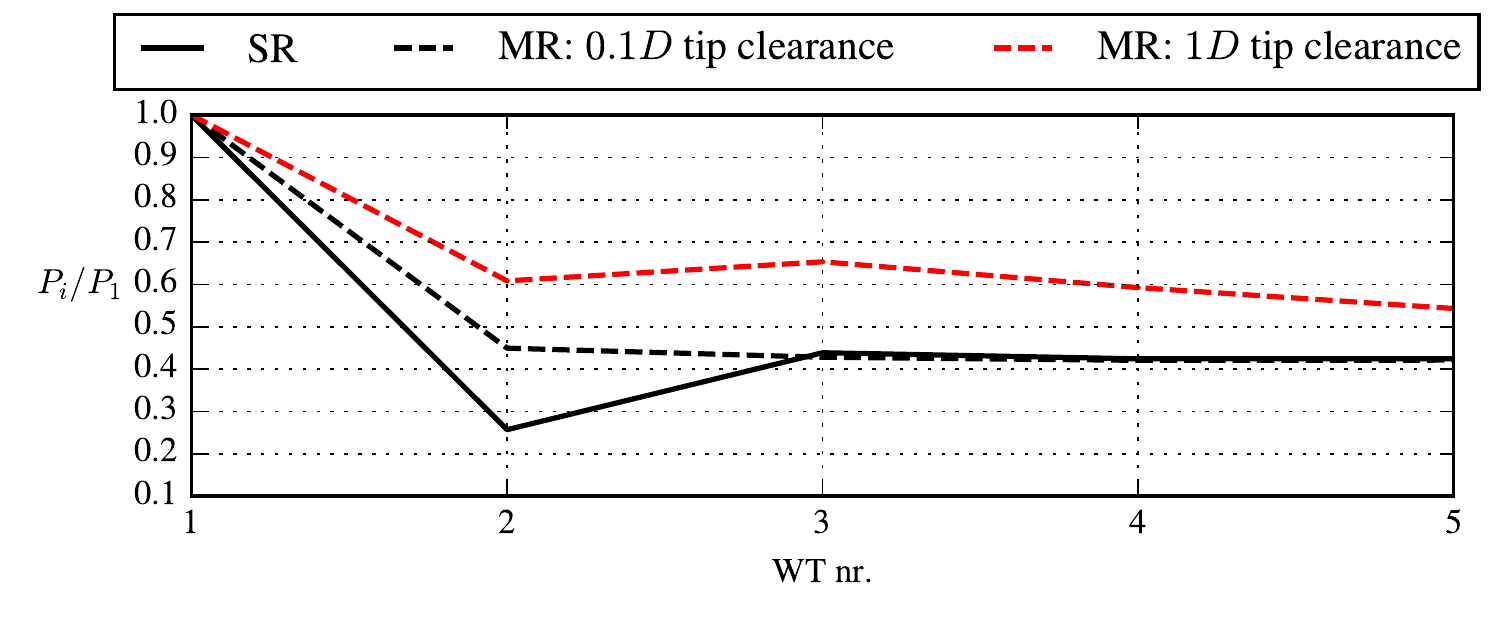}
\caption{LES results of the wind turbine (WT) power in a row of five wind turbines, normalized by the power of the first wind turbine. Results are shown for SR and MR wind turbines for two different tip clearances, for a wind speed of 8 m/s and an aligned wind direction of 270\degree.}
\label{fig:WFrowLEScase}
\end{figure}
Figure~\ref{fig:WFrowLEScase}, shows that the power deficit of a SR and MR wind turbine row is only significantly different for the first downstream wind turbine (using a tip clearance of $0.1D$), as also observed in the RANS simulations using a tip clearance of $0.05D$.
Note that changing the tip clearance from $0.1D$ to $0.05D$ has a negligible effect on the results.
However, when the tip clearance between the rotors of the MR wind turbine is set to $1D$, the difference in power deficit between the SR and MR wind turbine row increases and is still present for the entire row, which explains the LES results of Ghasais et al. \cite{Ghaisas18}.

\subsubsection{AEP}\label{sec:ResAEP}
\begin{figure}[!h]
\centering
\includegraphics[scale=0.8,clip=true,trim=0 10 0 0]{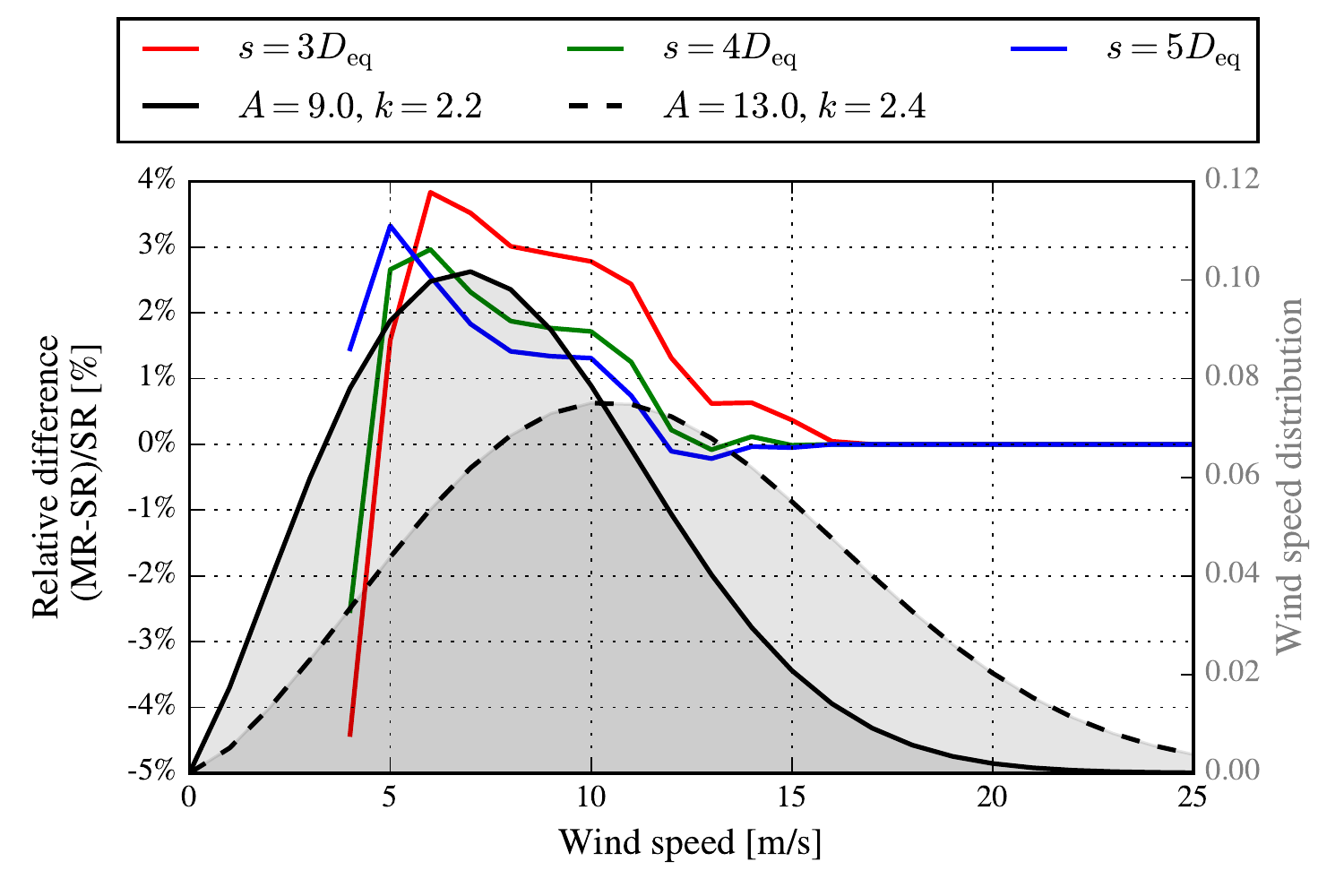}
\caption{Relative difference in wind farm power integrated over all wind directions between the SR and MR wind farms, for different inter spacings. Two Weibull wind speed distributions are shown that are used to compute the difference in AEP, as listed in Table \ref{tab:AEPdiff}.}
\label{fig:AEPdiff}
\end{figure}

The relative difference between the SR and MR wind farms in terms of wind farm efficiency per wind speed, summed over all wind directions is depicted in Figure~\ref{fig:AEPdiff}, for the three different inter spacings.
The largest relative difference of about 4\% is simulated for the smallest inter spacing of $3D_{\rm eq}$ and below rated, where the thrust coefficient is the highest, as depicted in Figure~\ref{fig:Pct}.
For low wind freestream speeds of 4 and 5 m/s, the SR wind farm is performing better than the MR wind farm for the two smallest inter spacings, which is related to the wind shear.
There are low wind speed inflow cases where a downstream SR wind turbine (from the SR wind farm) is just operating above the cut-in wind speed.
When the same inflow case is occurring in the MR windfarm, the bottom MR wind turbines will see a wind speed below the cut-in wind speed and are shut down, while the top rotor are just in operation because the wind speed is larger at $z=h_2$.
The better performance of the SR wind farm for low wind speed is not very relevant for the AEP comparison between SR and MR wind farms because most energy is gained for the higher wind speed regime.

The difference in AEP between the SR and MR wind farms is calculated in Table \ref{tab:AEPdiff} with two different Weibull distributions (also plotted in Figure~\ref{fig:AEPdiff}); 1) Offshore meteorological mast measurements located at the Horns Rev 2 offshore wind farm, at a height between 15 and 20 m \cite{Hansen12}: $A=9.0$ m/s  $k=2.2$, and 2) Offshore lidar measurements taken at a height of 122 m located at a German research platform in the North Sea (FINO3) \cite{Gryning16}: $A=13$ m/s and $k=2.4$.
A uniform distributed wind rose is used.
Table \ref{tab:AEPdiff} shows that the MR wind farm produces up to 1.7\% more AEP than the SR wind farm, which is calculated for the smallest inter spacing of $3D_{\rm eq}$ and using the Weibull distribution with the lower mean wind speed ($A=9.0$ m/s, $k=2.2$). 
When the other Weibull distribution is employed ($A=13.0$ m/s, $k=2.4$), the AEP gain of the MR wind farm with the smallest inter spacing is about 1\%.
This shows that the choice of the Weibull distribution has a large impact on quantifying the AEP difference between the SR and MR wind farms.
For the larger inter spacings, the gain in AEP for the MR wind farm decreases to 0.3\% and 0.7\% for the wind speed distributions with the lower and higher mean wind speeds, respectively.
The calculated gain in AEP is only a result of the faster wake recovery of a MR wind turbine wake.
The AEP gain of a MR wind turbine due to the rotor interaction, as simulated in previous work \cite{Laan19}, has not been included in the simulations.
Hence, the total AEP gain of a MR wind farm is expected to be larger than presented in Table \ref{tab:AEPdiff}.

According to the RANS simulations, the main gain in AEP for the MR wind farm is achieved for the aligned wind directions and the first downstream wind turbines.
Hence, one could think of a wind farm layout where most of the wake interaction involves two wind turbines, which could result in a larger gain in AEP when using MR wind turbines instead of SR wind turbines.

\begin{table}[!h]
    \centering
    \begin{tabular}{l|ccc}
    Inter wind turbine spacing [$D_{\rm eq}$]    &  3 & 4 & 5\\
    \hline
    AEP gain MR wind farm ($A=9.0$ m/s and $k=2.2$) [\%] & 1.7 & 0.95 & 0.69\\
    AEP gain MR wind farm ($A=13$ m/s and $k=2.4$) [\%]    & 0.97 &  0.46 & 0.30
    \end{tabular}
    \caption{AEP gain of the MR wind farm due to the faster wake recovery, for two different Weibull distributions with $A$ as the scaling parameter and $k$ as the shape parameter, and three inter spacings.}
    \label{tab:AEPdiff}
\end{table}

\section{Conclusions}
The difference in wind turbine wake interaction between single-rotor (SR) and multi-rotor (MR) wind farms is investigated with Reynolds-averaged Navier-Stokes simulations using a $4\times4$ rectangular wind farm layout for three different inter spacings: 3, 4 and $5D_{\rm eq}$, where $D_{\rm eq}$ is the rotor diameter of the SR wind turbine.
The MR wind farm simulations predict 0.3-1.7\% more annual energy production (AEP) compared to the SR wind farm simulations, where the highest gain is obtained for the tightest inter wind turbine spacing of $3D_{\rm eq}$.
The gain in AEP is mainly caused by the aligned wind directions for the first downstream wind turbine in a wind turbine row of the MR wind farm, which experiences 7-16\% more power compared to the same wind turbine in SR wind farm due to the faster wake recovery of the MR wind turbine wake.
Further downstream, the wind turbines in the SR and MR wind farms produce a similar power mainly because of the enhanced mixing due to wake turbulence of the upstream wind turbines, and a difference in thrust force for the first downstream wind turbine.
The RANS results are in contradiction with results from large-eddy simulations (LES) performed by Ghaisas et al. \cite{Ghaisas18}, who showed that all downstream MR wind turbines in a row of five wind turbines produce 15-45\% more power compared to a row of SR wind turbines.
In this work, an additional LES test case is performed to show that the results of Ghaisas et al. \cite{Ghaisas18} are caused by an unrealistically large tip clearance of one small rotor diameter $D$.
For a more common tip clearance of $0.1D$, our LES predicts that only the first downstream MR wind turbine in a row of five wind turbines produces more power compared to a row of SR wind turbines, which verifies the RANS results.
In addition, the RANS simulations show that wind farm efficiency of the SR and MR wind farms is similar for non-aligned wind directions.
The gain in AEP of a MR wind farm could be further increased by using alternative wind farm layouts, where most wake interactions involve only two wind turbines, which should be investigated in future work.

\section*{References}
\bibliographystyle{iopart-num} 
\bibliography{bibliographyIOP}

\providecommand{\newblock}{}
\begin{thebibliography}{10}
\expandafter\ifx\csname url\endcsname\relax
  \def\url#1{{\tt #1}}\fi
\expandafter\ifx\csname urlprefix\endcsname\relax\def\urlprefix{URL }\fi
\providecommand{\eprint}[2][]{\url{#2}}

\bibitem{Jamieson14}
Jamieson P, Chaviaropoulos T, Voutsinas S, Branney M, Sieros G and
  Chasapogiannis P 2014 {\em PO.ID 203 EWEC \& Excibition Barcelona, EWEA\/}
  p~1

\bibitem{Laan19}
van~der Laan M~P, J A~S, Ramos~Garc\'ia N, Angelou N, Pirrung G~R, Ott S,
  Sj\"{o}holm M, S{\o}rensen K~H, Vianna~Neto J~X, Kelly M, K M~T and Larsen
  G~C 2019 {\em Wind Energy Science Discussions\/}  1

\bibitem{Ghaisas18}
Ghaisas N~S, Ghate A~S and Lele S~K 2018 {\em Journal of Physics: Conference
  Series\/} {\bf 1037} 1

\bibitem{DTU10MW}
Bak C, Zahle F, Bitsche R, Kim T, Yde A, Henriksen L~C, Natarajan A and Hansen
  M~H 2013 Description of the dtu 10 mw reference wind turbine Tech. Rep.
  I-0092 Technical University of Denmark

\bibitem{MikkelsenRPhD}
Mikkelsen R 2003 {\em Actuator {D}isc {M}ethods {A}pplied to {W}ind
  {T}urbines\/} Ph.D. thesis DTU

\bibitem{ADval}
R\'ethor\'e P~E, van~der Laan M~P, Troldborg N, Zahle F and S{\o}rensen N~N
  2013 {\em Wind Energy\/} Pub. online

\bibitem{Laan14c}
van~der Laan M~P, S{\o}rensen N~N, R{\'e}thor{\'e} P~E, Mann J, Kelly M~C,
  Troldborg N, Hansen K~S and Murcia J~P 2015 {\em Wind Energy\/} {\bf 18} 2065

\bibitem{Laan13b}
van~der Laan M~P, S{\o}rensen N~N, R{\'e}thor{\'e} P~E, Mann J, Kelly M~C,
  Troldborg N, Schepers J~G and Machefaux E 2015 {\em Wind Energy\/} {\bf 18}
  889

\bibitem{Laan14b}
van~der Laan M~P, S{\o}rensen N~N, R{\'e}thor{\'e} P~E, Mann J, Kelly M~C and
  Troldborg N 2015 {\em Wind Energy\/} {\bf 18} 2223

\bibitem{NielsPHD}
S{\o}rensen N~N 1994 {\em General purpose flow solver applied to flow over
  hills\/} Ph.D. thesis DTU

\bibitem{Basis3D}
Michelsen J~A 1992 Basis3d - a platform for development of multiblock {PDE}
  solvers. Tech. rep. DTU

\bibitem{QUICK}
Leonard B~P 1979 {\em Computer Methods in Applied Mechanics and Engineering\/}
  {\bf 19} 59

\bibitem{SIMPLE}
Patankar S~V and Spalding D~B 1972 {\em International Journal of Heat and Mass
  Transfer\/} {\bf 15} 1787

\bibitem{RhieChowMod}
R\'ethor\'e P~E and S{\o}rensen N~N 2012 {\em Wind Energy\/} {\bf 15} 915

\bibitem{ADNielsRev}
Troldborg N, S{\o}rensen N~N, R\'ethor\'e P~E and van~der Laan M~P 2015 {\em
  Computers and Fluids\/} {\bf 119} 197

\bibitem{nsqrBC}
S{\o}rensen N~N, Bechmann A, Johansen J, Myllerup L, Botha P, Vinther S and
  Nielsen B~S 2007 {\em Journal of Physics: Conference series\/} {\bf 75} 1

\bibitem{Laan18a}
van~der Laan M~P and Andersen S~J 2018 {\em Journal of Physics: Conference
  Series\/} {\bf 1037} 1

\bibitem{Pirrung18}
Pirrung G~R and van~der Laan M~P 2018 {\em Wind Energy\/} To be submitted

\bibitem{mixedOrder}
Roy C~J 2003 {\em American Institute of Aeronautics and Astronautics Journal\/}
  {\bf 41}

\bibitem{Hansen12}
Hansen K~S, Barthelmie R~J, Jensen L~E and Sommer A 2012 {\em Wind Energy\/}
  {\bf 15} 183

\bibitem{Gryning16}
Gryning S~E, Floors R, Pe\~{n}a A, Bathvarova E and Br\"{u}mmer B 2016 {\em
  Boundary-Layer Meteorol\/} {\bf 159} 329

\bibitem{Porte-Agel2011}
Port{\'e}-Agel F, Wu Y~T, Lu H and Conzemius R~J 2011 {\em J. Wind Eng. Ind.
  Aerodyn.\/} {\bf 99} 154--168

\bibitem{Abkar2013}
Abkar M and Port{\'e}-Agel F 2013 {\em Energies\/} {\bf 6} 2338--2361

\bibitem{Abkar2015a}
Abkar M and Port{\'e}-Agel F 2015 {\em Phys. Fluids\/} {\bf 27} 035104

\bibitem{Abkar2016b}
Abkar M and Port{\'e}-Agel F 2016 {\em Phys. Rev. Fluids\/} {\bf 1}(6) 063701

\bibitem{Yang2018}
Yang X~I and Abkar M 2018 {\em J. Fluid Mech.\/} {\bf 842} 354--380

\bibitem{Abkar2016}
Abkar M, Sharifi A and Port{\'e}-Agel F 2016 {\em J. Turbul.\/} {\bf 17}
  420--441

\end{thebibliography}

\appendix
\section{Methodology of additional large-eddy simulations}\label{sec:AdditionalLEScase}
The LES code used is this study solves the filtered mass conservation and momentum equations. The scale-dependent Lagrangian dynamic model is employed for the subgrid-scale modeling, and the turbine effect is accounted via the standard actuator-disk approach. 
Detailed description of the numerical code can be found in  \cite{Porte-Agel2011,Abkar2013,Abkar2015a,Abkar2016b,Yang2018}.  
To generate the inflow condition, a precursor simulation of the neutrally stratified boundary-layer flow is performed. 
The computational domain is $3200\times800\times355$ m$^3$ in the $x$, $y$, and $z$ directions, respectively, and it is divided uniformly into $320\times160\times72$ grid points.  
The boundary condition is periodic in the lateral direction, and a fringe zone is applied in the streamwise direction to smoothly adjust the flow from the very-far-wake downwind condition to that of the precursor simulation \cite{Abkar2016}. 
The wind farm consists of 10 wind turbines arranged in two rows of five wind turbines with 5$D_{eq}$ inter spacing. The thrust coefficient of the turbines is set to $0.75$. The diameters of the SR and MR turbines are $D_{eq}=80$ m and $D=40$ m, respectively. The SR hub height and the center of the MR turbines are located at $70$ m. 
The inflow has a mean hub-height velocity of about 8 m/s with a total turbulence level of $5.4$\%. 

\end{document}